\documentclass[11pt]{article}
\usepackage{moriond,epsfig}
\input{epsf}
\usepackage{graphicx}

\def\Journal#1#2#3#4{{#1} {\bf #2}, #3 (#4)}

\def\PLB{{\em Phys. Lett.}  B}
\def\PRL{\em Phys. Rev. Lett.}
\def\PRD{{\em Phys. Rev.} D}

\def\be{\begin{equation}}
\def\ee{\end{equation}}
\def\bea{\begin{eqnarray}}
\def\eea{\end{eqnarray}}

\begin{document}
\vspace*{4cm}
\title{RECENT BOTTOMONIUM AND CHARMONIUM RESULTS FROM CLEO}

\author{ T.~FERGUSON }

\address{Department of Physics, Carnegie Mellon University,\\
Pittsburgh, PA 15213 USA}

\maketitle\abstracts{Recent results from the CLEO experiment on decays
and partial widths of various bottomonium and charmonium resonances
are presented. New measurements of $\Gamma_{ee}$ for the 3 bound-state
Upsilons and the $J/\psi$ are discussed.  A determination of
$\sigma(\psi(3770) \rightarrow hadrons)$ is made, which solves a
20-year-old puzzle. The $Y(4260)$ state is confirmed and its decay
into 2 new modes is measured.}

\section{CESR and CLEO}

The Cornell electron-positron storage ring (CESR) is a symmetric
$e^+e^-$ collider located on the Cornell University campus in Ithaca,
NY. For almost 30 years, the accelerator has been running in the
Upsilon energy region from $\sqrt{s}$ = 9-11~$GeV$, providing the
accompanying detector, CLEO, with data to produce results on $B$,
$\Upsilon$, $\tau$, charm, and 2-photon physics.  Roughly 2 years ago,
the accelerator was modified with the addition of wiggler magnets to
allow it to run in the $\sqrt{s}$ = 3-5~$GeV$ charmonium energy
region.

CLEO~\cite{detector} is a standard $4\pi$ $e^+e^-$ magnetic
spectrometer.  In its latest configuration, called CLEO-c, it consists
of inner and main drift chambers for charged particle tracking, a CsI
electromagnetic calorimeter, and a Ring-Imaging Cherenkov detector
(RICH), all enclosed in a 1~T solenoidal magnetic field.  The charged
particle momentum resolution is 0.6\% at 1.0~$GeV$/c and the photon
energy resolution is 4\% at 100~$MeV$.  Paticle identification is done
using both the RICH and dE/dx measurements in the drift chambers.
Muon chambers interspersed in the magnet iron outside the solenoid
complete the detector.

\section{$\Gamma_{ee}$ of the $\Upsilon(1S, 2S, 3S)$}

The di-electron partial width, $\Gamma_{ee}$, is one of the basic
parameters of any heavy-quark bound system.  It is proportional to the
square of the state's wave function at the origin, and is a number
that all heavy-quark theories try to predict.  Furthermore, its
measurement can provide a stringent test of lattice QCD calculation.
However, the present world average precision of $\Gamma_{ee}$ for the
3 bound-state $\Upsilon$ resonances~\cite{PDG} is only 2.2\%, 4.2\%,
and 9.4\%, respectively.  To rectify this situation, CLEO performed
scans over the 3 resonances, as well as running below each resonance
to constrain the backgrounds.

CLEO uses a standard procedure to measure $\Gamma_{ee}$ -- the total
hadronic cross section is measured over each resonance.  The integral
of this cross section with respect to the center-of-mass energy is
proportional to $\Gamma_{ee} \Gamma_{had} / \Gamma_{tot}$, where
$\Gamma_{had}$ and $\Gamma_{tot}$ are the resonance's hadronic and
total widths, respectively.  If we then assume lepton universality, 
taking the 3 leptonic branching ratios as equal ($B_{ee} = B_{\mu\mu}
= B_{\tau\tau}$), we can solve for $\Gamma_{ee}$ using:

\begin{equation}
\Gamma_{ee} \;=\; \frac{\Gamma_{ee} \Gamma_{had}}{\Gamma_{tot} (1 -
3B_{\mu\mu})}.
\label{eq:gamma}
\end{equation}

The main backgrounds include the continuum $e^+e^- \rightarrow
hadrons$ process with a $1/s$ energy dependence, two-photon production
($e^+e^- \rightarrow e^+e^-X$) with a $ln(s)$ dependence, cosmic rays,
beam-gas interactions, and events from the high-energy tails of the
$\Upsilon(1S)$ and $\Upsilon(2S)$.  The beam-gas and cosmic ray
backgrounds are subtracted using data from special single- and no-beam
runs.

The hadronic cross section measurements for each resonance are then
fit to a convolution of a Breit-Wigner function, including
interference between the resonance and the continuum hadronic
production, initial-state radiation, a Gaussian for the 4~MeV CESR
beam energy spread, and background terms proportional to $1/s$ and
$ln(s)$. The resulting statistical errors on the partial widths are
0.3\% (1S), 0.7\% (2S), and 1.0\% (3S).  The main systematic errors
are from uncertainties in the luminosity measurement (1.3\%) and the
hadronic event efficiency (0.5\%).  The measured parameters, including
the total width of each resonance, found from $\Gamma_{ee}$ by
assuming that $B_{ee}$ = $B_{\mu\mu}$ and using a recent, very precise
CLEO measurement of $B_{\mu\mu}$~\cite{1Smumu}, are given in
Table~\ref{tab:widths}.

\begin{table}[htbp]
\caption{CLEO measurements of various resonance parameters for the 3
  bound-state Upsilons. The first errors are statistical, the second
  are systematic, and the third for $\Gamma_{tot}$ are due to the
  uncertainty on $B_{\mu\mu}$. \label{tab:widths}}
\vspace{0.4cm}
\begin{center}
\begin{tabular}{|c|c|c|c|}
\hline
& & &  \\
& $\Upsilon(1S)$ & $\Upsilon(2S)$ & $\Upsilon(3S)$ \\ \hline
& & &  \\
$\Gamma_{ee} \Gamma_{had} / \Gamma_{tot} \; (keV)$   &  
$ 1.252 \pm 0.004 \pm 0.019$ & 
$ 0.581 \pm 0.004 \pm 0.009$ &
$ 0.413 \pm 0.004 \pm 0.006$ \\
$\Gamma_{ee} \; (keV)$ &
$ 1.354 \pm 0.004 \pm 0.020$ &
$ 0.619 \pm 0.004 \pm 0.010$ &
$ 0.446 \pm 0.004 \pm 0.007$ \\
$\Gamma_{tot} \; (keV)$ &
$ 54.4 \pm 0.2 \pm 0.8 \pm 1.6$ &
$ 30.5 \pm 0.2 \pm 0.5 \pm 1.3$ &
$ 18.6 \pm 0.2 \pm 0.3 \pm 0.9$ \\
& & & \\
& $\Upsilon(2S)/\Upsilon(1S)$ & $\Upsilon(3S)/\Upsilon(1S)$ &
$\Upsilon(3S)/\Upsilon(2S)$ \\ \hline 
& & & \\
$\Gamma_{ee}(mS) / \Gamma_{ee}(nS)$ &
$ 0.457 \pm 0.004 \pm 0.004$ &
$ 0.329 \pm 0.003 \pm 0.003$ &
$ 0.720 \pm 0.009 \pm 0.007$ \\
& & &  \\ \hline
\end{tabular}
\end{center}
\end{table}

These new measurements of $\Gamma_{ee}$~\cite{123ee} have improved on
the precision of the previous world averages~\cite{PDG} by factors of
1.5 (1S), 2.5 (2S) and 5.2 (3S).  To compare these results to the
latest unquenched lattice QCD calculations~\cite{QCD}, we use the
combination:

\begin{equation}
\frac{\Gamma_{ee}(2S) \; M^2(2S)}{\Gamma_{ee}(1S) \; M^2(1S)}.
\end{equation}

Our measured value of $0.517 \pm 0.007$ for this variable is in good
agreement with the lattice QCD result~\cite{QCD}, extrapolated to zero
lattice spacing, of $0.48 \pm 0.05$. The hope is that the lattice QCD
calculations will eventually reach a precision of a few percent for
this variable and about a 10\% precision for $\Gamma_{ee}$ itself.
When this goal is achieved, the experimental measurements will now
have the precision needed for a meaningful comparison.

\section{$\Gamma_{ee}$ and $\Gamma_{tot}$ of the $J/\psi$}

Switching now to the charmonium sector, CLEO has made a new similar
measurement of $\Gamma_{ee}$ and $\Gamma_{tot}$ for the
$J/\psi$~\cite{psiee}. However, in this case the measurement was not
done by scanning over the resonance.  Instead, using data taken on the
$\psi(3770)$, we look for so-called radiative return events to the
$J/\psi$, in which an initial-state photon is radiated, followed by
the decay of the $J/\psi$ into $\mu^+\mu^-$.  After selecting di-muon
events with $M(\mu^+\mu^-) = M(J/\psi)$ and subtracting background
from radiative return to the $\psi(2S)$ and QED processes, the
resulting signal cross section is proportional to $B_{\mu\mu} \times
\Gamma_{ee}(J/\psi)$.  Dividing this value by the new CLEO measurement
of $B_{\mu\mu}(J/\psi)$~\cite{psimumu} (1.2\% precision), we obtain
$\Gamma_{ee}$ for the $J/\psi$.  Assuming lepton universality, $B_{ee}
= B_{\mu\mu}$, and dividing by $B_{\mu\mu}$ again, then gives
$\Gamma_{tot}(J/\psi)$.  The results are as follows:
\[ B_{\mu\mu} \times \Gamma_{ee}(J/\psi) \;=\; 0.3384 \pm 0.0058 \pm
  0.0071 \;keV, \]
\[\Gamma_{ee}(J/\psi) \;=\; 5.68 \pm 0.11 \pm 0.13 \; keV, \]
\[\Gamma_{tot}(J/\psi) \;=\; 95.5 \pm 2.4 \pm 2.4 \; keV, \]
where the first errors are statistical and the second are systematic.
Using a recent CLEO measurement of $\Gamma_{ee}(\psi(2S))$~\cite{2See}
found from an identical technique, we determine the ratio:
\[ \Gamma_{ee}(\psi(2S))/\Gamma_{ee}(J/\psi) \;=\; 0.45 \pm 0.01 \pm
0.02, \]
in which many of the systematic errors cancel.  All of these
measurements are more precise than the previous world average
values~\cite{PDG}.

\section{$\sigma(\psi(3770) \rightarrow hadrons)$ and
  $\Gamma_{ee}(\psi(3770))$} 

The $\psi(3770)$, the first $c\overline{c}$ resonance above open-charm
threshold, has had a long-standing puzzle about it.  The Lead-Glass
Wall~\cite{LGW} (1977) and Mark~II~\cite{MkII} (1981) experiments
first measured the total hadronic cross section for the $\psi(3770)$
to be $11.6 \pm 1.8$~nb.  Later, the Mark~III experiment~\cite{MkIII}
(1988), using a double-tag technique, found $\sigma(\psi(3770)
\rightarrow D\overline{D})$ = $5.0 \pm 0.5$~nb.  The large difference
between these 2 numbers was a complete surprise and has remained a
mystery, since it was believed that above open-charm threshold,
$\sigma(\psi(3770) \rightarrow non-D\overline{D}) << \sigma(\psi(3770)
\rightarrow D\overline{D})$.

CLEO recently repeated the Mark~III measurement, using an identical
technique but with much higher statistics (see the write-up by A.~Ryd
in these proceedings).  They found~\cite{DDcross}:
\[ \sigma(\psi(3770) \rightarrow D\overline{D}) \;=\; 6.39 \pm 0.10 \;
^{+0.17}_{-0.08}\; nb, \]
which is significantly higher than the Mark~III number, but still a
long way from the total hadronic cross section value. Thus, it
remained to also repeat the measurement of the total hadronic cross
section.  CLEO measures this cross section using the standard formula:
\begin{equation}
\sigma(\psi(3770) \rightarrow hadrons) \;=\;
\frac{N_{\psi(3770)}}{\epsilon_h {\cal L}},
\end{equation}
where $N_{\psi(3770)}$ is the observed number of hadronic decays of
the $\psi(3770)$, $\epsilon_h$ is the hadronic event efficiency
(80\%), and ${\cal L}$ is the total integrated luminosity ($281.3 \pm
2.8\;pb^{-1}$). To obtain $N_{\psi(3770)}$, we take the total number
of observed hadronic events and subtract off the number of events from
the continuum process $e^+e^- \rightarrow hadrons$, from the tails of
the $J/\psi$ and $\psi(2S)$, and from di-lepton events (especially
$\tau^+\tau^-$) faking hadrons.  The resulting cross
section~\cite{totalcross} is:
\[ \sigma(\psi(3770) \rightarrow hadrons) \;=\; 6.38 \pm 0.08
\; ^{+0.41}_{-0.30}\;nb, \]
where the first error is statistical and the second is systematic.
Subtracting CLEO's $D\overline{D}$ cross section from this gives:
\[ \sigma(\psi(3770) \rightarrow hadrons) \; - \; \sigma(\psi(3770)
\rightarrow D\overline{D}) \;=\; -0.01 \pm 0.08 \; ^{+0.041}_{-0.030} \;
nb. \]
This is consistent with a small non-$D\overline{D}$ branching fraction
for the $\psi(3770)$, which has been observed~\cite{2See,nonDD}, but
is much more in line with what was expected theoretically.  Thus, the
almost 20-year-old puzzle about the $\psi(3770)$ has been solved.

Using the $\sigma(\psi(3770) \rightarrow hadrons)$ measurement and
values for the mass and total width of the $\psi(3770)$~\cite{PDG}, we
can determine the di-electron partial width:
\[ \Gamma_{ee}(\psi(3770)) \;=\; 0.204 \pm 0.003 \;
^{+0.041}_{-0.027}\;keV. \]
This measurement is in good agreement and as precise as the PDG
value~\cite{PDG} of $0.26 \pm 0.04$, but has quite different
sources of systematic error.

\section{Charmonium Decays of the $\psi(4040)$, $\psi(4160)$, and
  $Y(4260)$}

Besides the $\psi(3770)$ discussed in the last section, prominent
structures in the $e^+e^-$ total hadronic cross section above
open-charm threshold are associated with the $\psi(4040)$,
$\psi(4160)$, and $\psi(4415)$ resonances.  All of these particles can
be assigned to specific $c\overline{c}$ states in both
non-relativistic and relativistic heavy-quark potential models.  They
are all characterized by large total widths, weaker couplings to
leptons than the bound-state resonances, and with predominant decays
to open charm.  Over the last few years, though, a number of new
states with masses around 4~GeV and with either open- or closed-charm
decays have been discovered.  The assignments of these particles in
the potential model schemes are not so obvious.  The latest of these
is the $Y(4260)$, discovered by BaBar~\cite{BaBar} through its decay
to $\pi^+\pi^- J/\psi$, though not yet confirmed by other experiments.
With a mass of 4259~MeV and a width of around 90~MeV, this particle
does not easily fit into any of the potential models, nor is its large
decay to a closed-charm channel what would be expected of a normal
$c\overline{c}$ state of this mass.  Furthermore, exactly at this
energy the $e^+e^-$ total hadronic cross section goes through a local
minimum -- hardly the normal signal for a $c\overline{c}$ resonance!

Thus, there have been a host of theoretical explanations for this new
resonance, including a ($c\overline{c}g$) hybrid charmonium
state~\cite{hybrid}, a $(cs)(\overline{c}\overline{s})$
tetraquark~\cite{tetra}, a $\chi_{cJ}\rho$~\cite{rho},
$\chi_{cJ}\omega$~\cite{omega}, or baryonium~\cite{bary} molecule, and
the normal $\psi(4S)\; c\overline{c}$ state, where interference
effects produce the dip in the open-charm cross section~\cite{4S}.
All of these theories have different predictions for the ratios of the
state's branching fractions to $\pi^+\pi^- J/\psi$,
$\pi^\circ\pi^\circ J/\psi$, and $K^+K^- J/\psi$.  For the $\psi(4S)$
possibility, the previously-assigned $\psi(4S)$ state, the
$\psi(4040)$, is expected to have an enhanced $\pi^+\pi^- J/\psi$
decay.

\begin{figure}[htbp]
\includegraphics[width=0.9\textwidth]{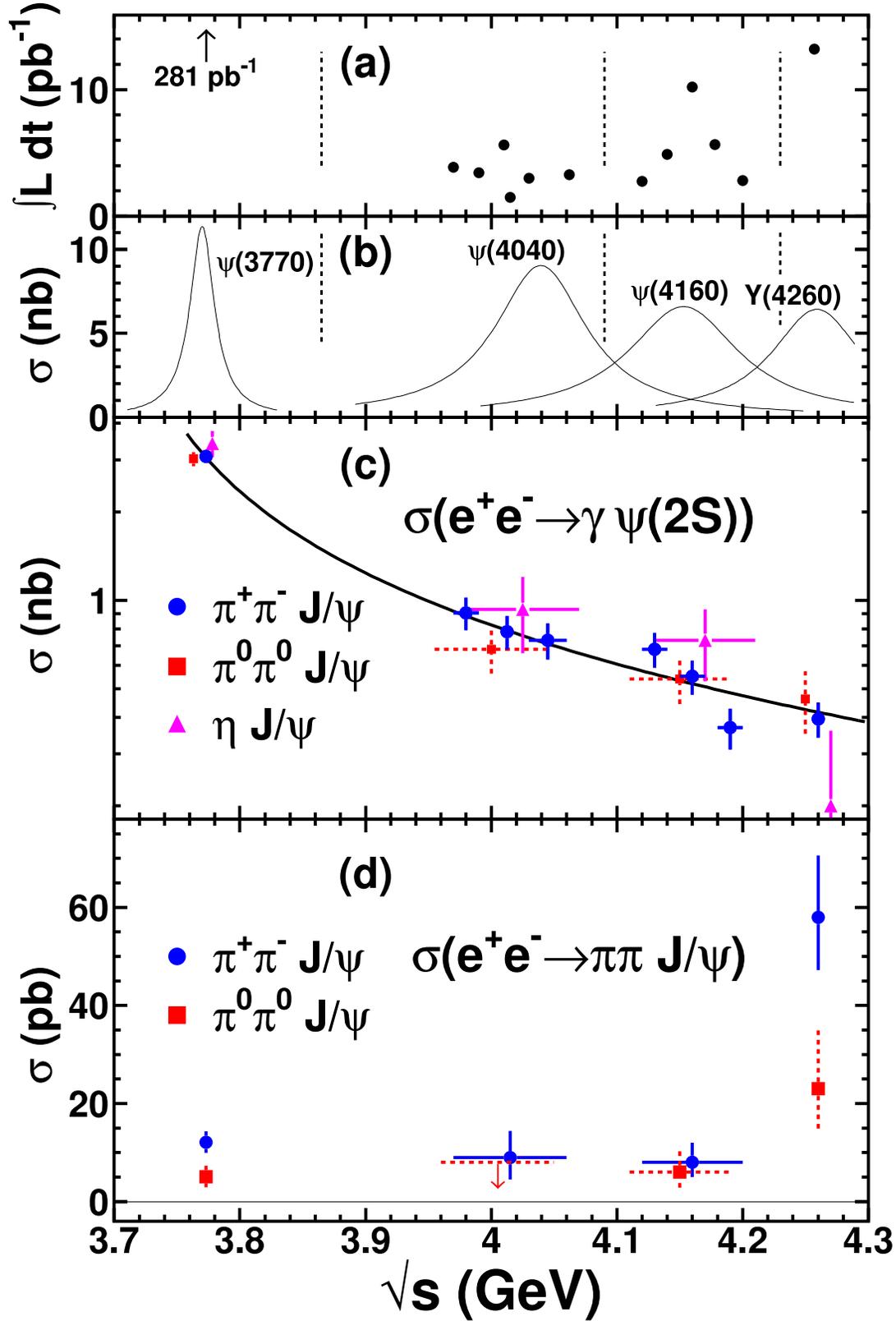}
\caption{(a)~Integrated luminosity versus $\sqrt{s}$ for the scan
  region.  (b)~The Breit-Wigner cross sections for the resonances in
  the scan energy region (the $Y(4260)$ vertical scale is arbitrary).
  The separation between the scan regions is shown by the vertical
  dotted lines.  (c)~Radiative-return cross section $e^+e^-
  \rightarrow \gamma \psi(2S)$ versus $\sqrt{s}$ for the 3 decay modes
  $\pi^+\pi^- J/\psi$ (circles), $\pi^\circ\pi^\circ J/\psi$ (squares
  and dashed lines), and $\eta J/\psi$ (triangles), along with the
  expected values (solid line) from the known $\psi(2S)$
  parameters. (d)~Cross sections for $e^+e^- \rightarrow \pi^+\pi^-
  J/\psi$ (circles) and $\pi^\circ\pi^\circ J/\psi$ (squares, dashed
  lines) versus $\sqrt{s}$.
\label{fig:cross-sections}}
\end{figure}

To try to confirm and clarify the $Y(4260)$, CLEO performed a scan
from $\sqrt{s}$ = 3.97~--~4.26~GeV, as well as using data taken at the
$\psi(3770)$, looking for decays to 16 different closed-charm final
states containing a $J/\psi$, $\psi(2S)$, $\chi_{cJ}$, or
$\phi$~\cite{scan}.  The integrated luminosity as a function of
$\sqrt{s}$ is shown in Fig.~\ref{fig:cross-sections}(a).  We break the
scan up into distinct regions, shown by the vertical dotted lines in
Fig.~\ref{fig:cross-sections}(b), according to the resonance each scan
point is nearest: $\psi(4040)$ for $\sqrt{s}$ = 3.97~--~4.06~GeV,
$\psi(4160)$ for $\sqrt{s}$ = 4.12~--~4.20~GeV, and $Y(4260)$ for
$\sqrt{s}$ = 4.26~GeV.

As a check on our analysis procedures and efficiencies, we first look
for the radiative-return process $e^+e^- \rightarrow \gamma \psi(2S)$,
with the $\psi(2S)$ then decaying into $\pi^+\pi^- J/\psi$,
$\pi^\circ\pi^\circ J/\psi$, or $\eta J/\psi$.  These are 3 of the 16
modes searched for in the scan.  Given the parameters of the
$\psi(2S)$ and its known branching ratios to these 3 modes, one can
predict the cross section for the radiative-return process as
a function of $\sqrt{s}$.  This prediction and our measured cross
sections are shown in Fig.~\ref{fig:cross-sections}(c).  There is
excellent agreement between our measurements and the expected values.

\begin{figure}[htbp]
\includegraphics[width=0.9\textwidth]{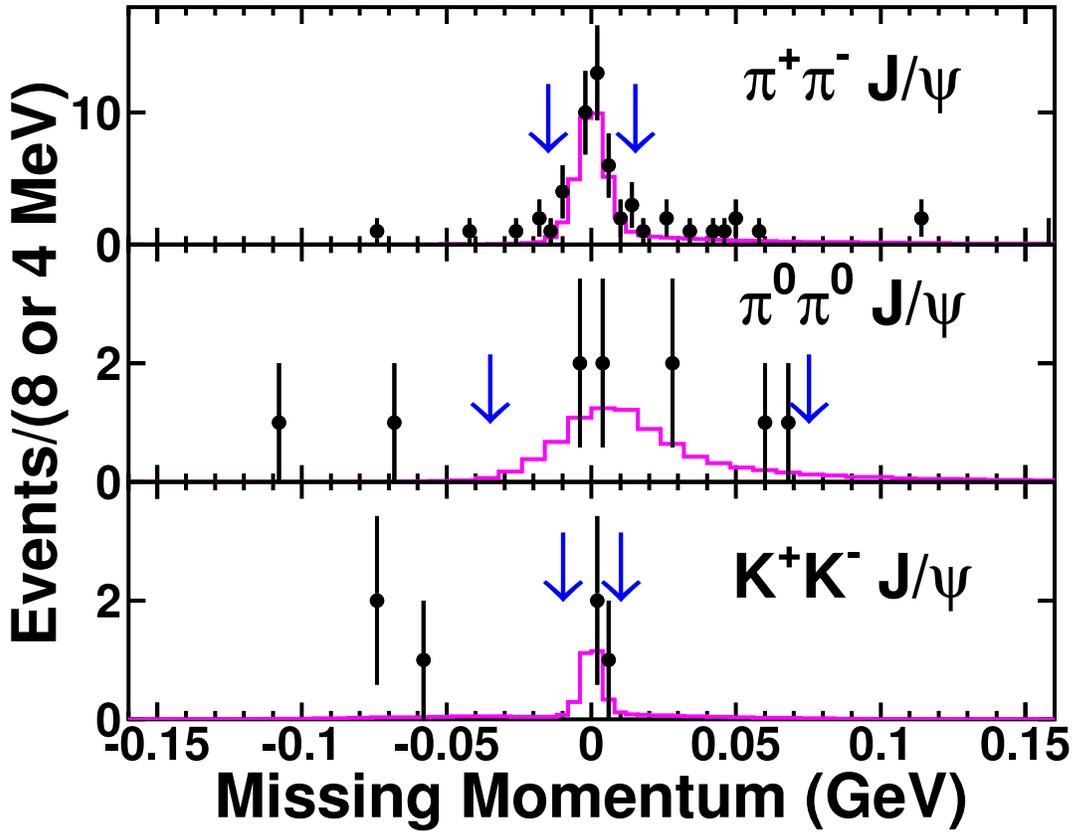}
\caption{The missing-momentum distributions for: (top) $\pi^+\pi^-
  J/\psi$, (middle) $\pi^\circ\pi^\circ J/\psi$, and (bottom) $K^+K^-
  J/\psi$ from the data at $\sqrt{s}$ = 4.26~GeV.  The signal shape
  expected from Monte Carlo simulations, scaled to the net signal
  size, is shown by the solid line histograms.  The cuts for each
  signal are shown by the arrows in each plot.
\label{fig:missing-momentum}}
\end{figure}

With our analysis technique validated, we now turn to the scan itself.
When any of the 16 decay modes is identified, a signal is searched for
by requiring that the magnitude of the total missing momentum in the
event should be consistent with 0.  We observe only 3 statistically
significant signals in the modes $\pi^+\pi^- J/\psi$ ($11\sigma$),
$\pi^\circ\pi^\circ J/\psi$ ($5.1\sigma$), and $K^+K^- J/\psi$
($3.7\sigma$), all from the $Y(4260)$ scan region.  Their corresponding
missing-momentum distributions are shown in
Fig.~\ref{fig:missing-momentum}. The resulting cross sections for the
$\pi^+\pi^- J/\psi$, and $\pi^\circ\pi^\circ J/\psi$ modes are shown
as a function of $\sqrt{s}$ in Fig.~\ref{fig:cross-sections}(d).  The
signal for the $Y(4260)$ is clear.

Thus, CLEO has confirmed the BaBar discovery of the $Y(4260)$ decaying
into $\pi^+\pi^- J/\psi$ and has made the first observation of its decay
into $\pi^\circ\pi^\circ J/\psi$ and $K^+K^- J/\psi$.  We find no
evidence for other decays in the 3 resonance regions.  In particular,
we set upper limits of:
\[ B(\psi(4040) \rightarrow \pi^+\pi^- J/\psi) \;<\; 0.4\%, \]
\[ B(\psi(4160) \rightarrow \pi^+\pi^- J/\psi) \;<\; 0.4\%. \]

The observation of the $\pi^\circ\pi^\circ J/\psi$ decay mode
disfavors the $\chi_{cJ}\rho$ molecular model~\cite{rho}.  The fact
that the $\pi^\circ\pi^\circ J/\psi$ rate is consistent with half of
the $\pi^+\pi^- J/\psi$ rate disagrees with the prediction of the
baryonium model~\cite{bary}, and the observation of the $K^+K^-
J/\psi$ mode is incompatible with both of these models.  The lack of
an enhancement for $\psi(4040) \rightarrow \pi^+\pi^- J/\psi$ makes
the identification of the $Y(4260)$ as the $\psi(4S)$~\cite{4S} less
attractive.  All of our results are compatible with the
hybrid-charmonium interpretation~\cite{hybrid} of the $Y(4260)$.
However, to completely rule-out or verify the remaining theories,
searches for the $Y(4260)$ decay to open-charm states will have to be
performed.  This will be quite difficult, though, given the very small
open-charm total cross section at that center-of-mass energy.

\end{document}